\documentclass[aps,prb,twocolumn,floatfix,showpacs,superscriptaddress,fleqn]{revtex4-1}
\usepackage{amssymb}

\usepackage{graphicx}
\usepackage{amsmath}
\usepackage{bm}
\usepackage{color}

\begin{document}

\author{A. A. \surname{Mitioglu}}
\affiliation{Laboratoire National des Champs Magn\'etiques Intenses, UPR 3228, CNRS-UJF-UPS-INSA, Grenoble and
Toulouse, France}\affiliation{Institute of Applied Physics, Academiei Str. 5, Chisinau, MD-2028, Republic of Moldova}

\author{P. \surname{Plochocka}}
\affiliation{Laboratoire National des Champs Magn\'etiques Intenses, UPR 3228, CNRS-UJF-UPS-INSA, Grenoble and
Toulouse, France}\email{paulina.plochocka@lncmi.cnrs.fr}

\author{J. N. \surname{Jadczak}}
\affiliation{Laboratoire National des Champs Magn\'etiques Intenses, UPR 3228, CNRS-UJF-UPS-INSA, Grenoble and
Toulouse, France}\affiliation{Institute of Physics, Wroclaw University of Technology, Wybrzeze Wyspianskiego 27,
50-370, Wroclaw, Poland}

\author{W. \surname{Escoffier}}
\affiliation{Laboratoire National des Champs Magn\'etiques Intenses, UPR 3228, CNRS-UJF-UPS-INSA, Grenoble and
Toulouse, France}

\author{G. L. J. A. \surname{Rikken}}
\affiliation{Laboratoire National des Champs Magn\'etiques Intenses, UPR 3228, CNRS-UJF-UPS-INSA, Grenoble and
Toulouse, France}

\author{L. \surname{Kulyuk}}
\affiliation{Institute of Applied Physics, Academiei Str. 5, Chisinau, MD-2028, Republic of Moldova}

\author{D. K. \surname{Maude}}
\affiliation{Laboratoire National des Champs Magn\'etiques Intenses, UPR 3228, CNRS-UJF-UPS-INSA, Grenoble and
Toulouse, France}

\title
{Optical manipulation of the exciton charge state in single layer tungsten disulfide}


\date{\today}

\begin{abstract}
Raman scattering and photoluminescence (PL) emission are used to investigate a single layer of tungsten disulfide
(WS$_{2}$) obtained by exfoliating n-type bulk crystals. Direct gap emission with both neutral and charged exciton
recombination is observed in the low temperature PL spectra. The ratio between the trion and exciton emission can be
tuned simply by varying the excitation power. Moreover, the intensity of the trion emission can be independently tuned
using additional sub band gap laser excitation.
\end{abstract}

\maketitle

\section{Introduction}

Layered compounds involving transition metals from group VI and chalcogens (the so-called dichalcogenides) are
promising candidates for exploring atomically thin structures. The basic building block consists of a monolayer of a
transition metal with a chalcogen monolayer above and below. We refer to this chalcogen-metal-chalcogen stack as a
single layer. All the dichalcogenides have a strong intra layer chalcogen-metal covalent bond while, the layers are
weakly held together by van der Waals forces. Nevertheless, the inter layer coupling plays a significant role in
determining the band structure. Bulk crystals are semiconductors with an indirect gap in the near infrared spectral
range. In contrast, single layer transition metal dichalcogenides such as molybdenum disulfide (MoS$_{2}$), tungsten
disulfide (WS$_{2}$) or tungsten diselenide (WSe$_{2}$) are two dimensional (2D) semiconductors with \emph{a direct gap
in the visible spectral range}.~\cite{Mak10,Splendiani10,Eda11,Albe02,Gutierrez13,Zhao13,Wang12,Cao12,Mak13} The
optical response of a single layer of these materials is dominated by excitonic effects; the optical spectrum is
characterized by the presence of two low-energy exciton  peaks (A and B excitons) that arise from vertical transitions
from a spin-orbit-split valence band to a doubly degenerate conduction band at the K point of the Brillouin
zone.~\cite{Frey98,Klein01,Ramasubramaniam12}

The physics of excitons in 2D semiconductors is known to be extremely rich once additional carriers are introduced into
the system; the optical spectra consist of emission from both neutral (X) and charged excitons (X$^{\pm}$). Charged
excitons were discovered in II-VI and III-V quantum wells (QWs).~\cite{Kheng93,Finkelstein95} In GaAs QWs it has been
shown that the ratio between neutral and charged exciton can be tuned by using a back
gate,~\cite{Groshaus07,Bar-Joseph05,Finkelstein95} or additional illumination.~\cite{Shields95,Glasberg99,Jadczak12} In
CdTe QWs this ratio can simply be tuned using additional illumination or a change of the excitation
power.~\cite{Huard00,Esser00,Kossacki99} Recently, it has been shown that a gate can be used to tune the carrier
density in single layer molybdenum disulfide and molybdenum diselenide.~\cite{Mak13,Ross13} While in standard
semiconductors the dissociation energy of the charged exciton is relatively small ($\simeq$ few meV), in exfoliated
dichalcogenides this energy is approximately an order of magnitude larger. Hence, the ability to control the exciton
charge state in semiconductor structures which emit light at room temperature and in the visible range, is expected to
open many possibilities for optoelectronics applications.

In this paper we show that in a single layer of tungsten disulfide (WS$_{2}$) obtained by the exfoliation of n-type
bulk crystals, we observe both charged and neutral exciton recombination in the photoluminescence (PL) emission
spectra. Additionally, by simply changing the intensity of the laser excitation,  we can tune the ratio between the
trion and exciton emission which demonstrates our ability to tune the density of 2D carriers with light. Moreover,
using additional sub band gap laser excitation, the trion emission intensity can be independently tuned. Finally, the
temperature dependence of the direct gap of a WS$_{2}$ single layer is shown to follow the usual behavior for a
semiconductor.

\section{Experiment}

\begin{figure}[]
\begin{center}
\includegraphics[width= 8.5cm]{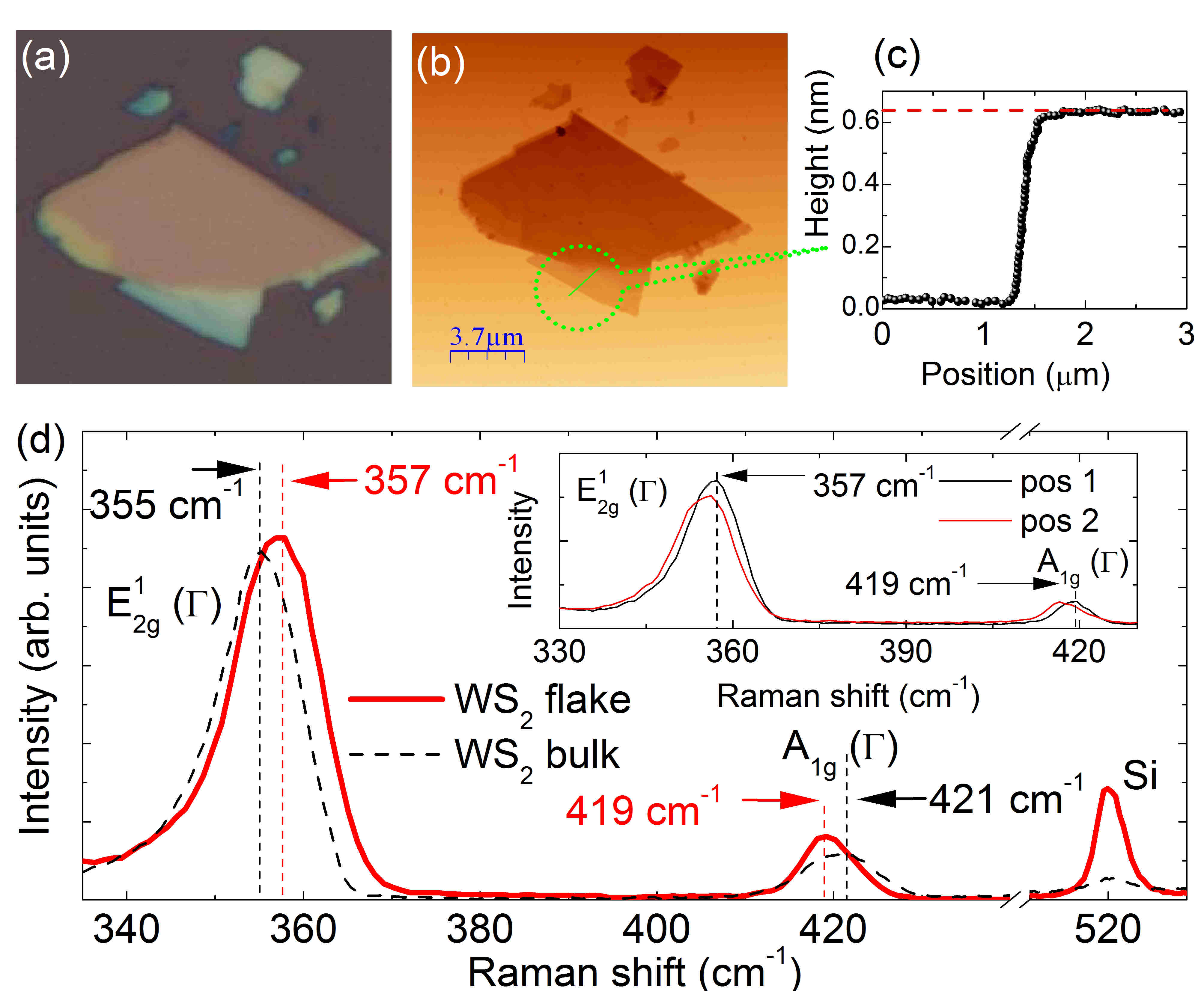}
\end{center}
\caption{(color online) (a) 0ptical microscope image and (b) AFM image of the WS$_2$ flake 1. (c) AFM height profile
along the path indicated by the green line (d) Raman spectra measured at T=300K on the single layer area of flake 1
(red line) and bulk WS$_2$ (black dashed line). The inset shows the strain induced shift of Raman spectra taken at two
different positions on the single layer area of flake 1.}\label{Fig1}
\end{figure}

Single and few layer flakes of tungsten disulfide (WS$_{2}$) have been obtained by mechanical exfoliation of bulk
2H-WS$_2$ (the hexagonal 2H-polytype of tungsten disulphide) single crystals grown using chemical vapor transport with
Bromine as the transport agent. Samples obtained in this way are naturally n-type.~\cite{Evans77} Hall measurements on
bulk crystals reveal an electron density $n_e \approx 10^{16}$~cm$^{-3}$ at room temperature, which decreases rapidly
at lower temperatures and is only $\approx 10^{12}$~cm$^{-3}$ at $T=140$K. After exfoliation, the WS$_{2}$ flakes were
placed on silicon substrate. Flakes with single layer regions have been identified using optical microscopy, atomic
force microscopy (AFM) and Raman spectroscopy. Typical results are presented in Fig~\ref{Fig1}. Parts of the flake
(subsequently referred to as flake 1), which are single layer have a characteristic blue color in the optical
microscope image and are also visible in the AFM image. The AFM height profile, measured moving along the green line
indicated on the AFM image, is shown Fig~\ref{Fig1}(c). The 0.6~nm step in the AFM height profile corresponds to a
single layer WS$_{2}$.\cite{Schutte87}

For the measurements the sample was placed in an optical cryostat mounted on motorized $x-y$ translation stages.
Excitation and collection was implemented using an external microscope objective giving a typical spot diameter of
$\simeq 1 \mu$m. The $\mu$PL and $\mu$-Raman spectra have been recorded using a spectrometer equipped with a CCD camera
and a laser emitting at $532$ nm was used for excitation. For some measurements the sample was simultaneously
illuminated by a Ti:Sapphire laser centered at 860~nm, with the laser beams entering collinearly to the microscope
objective.

\section{Micro-Raman and micro-PL}

Representative Raman spectra, obtained for a single layer region of flake 1(red line) and a bulk crystal (black dashed
line), are shown in Fig~\ref{Fig1}(d). The bulk crystal shows two Raman peaks at $355$ cm$^{-1}$ and $421 $cm$^{-1}$
corresponding to the well known active Raman modes $E_{2g}^{1}(\Gamma)$ and $E_{1g}(\Gamma)$. However, for the single
layer these two peaks shift towards each other by $2$ cm$^{-1}$. These results are consistent with previously reported
theoretical calculations and measurements on single layer of tungsten disulfide.~\cite{Albe02,Gutierrez13,Zhao13}
Hence, independently of the AFM measurements, the Raman data provides an additional confirmation of the single layer
character of the investigated regions of our flakes.

Fig.~\ref{Fig2} shows typical $\mu$PL spectra measured on a single layer region of two different flakes as a function
of the excitation power at $T=4$~K. Panel (a) corresponds to results obtained on flake 1. In both cases, at low
excitation powers, we observe strong emission around 2~eV which corresponds well with the predicted recombination of
the neutral exciton A across the direct gap of single layer WS$_{2}$.~\cite{Klein01} Moving around the flake the
measured emission energy can vary slightly around this value~\cite{Gutierrez13} which we attribute to strain as the
Raman spectra show simultaneous small rigid shift of the two Raman peaks with no change in the peak separation (see
inset of Fig~\ref{Fig1} d). This is consistent with the reported influence of the strain on the optical properties of a
single layer of MoS$_{2}$.~\cite{Conley13,Rice13}

\subsection {Neutral and charged exciton emission}

\begin{figure}[]
\begin{center}
\includegraphics[width= 8.5cm]{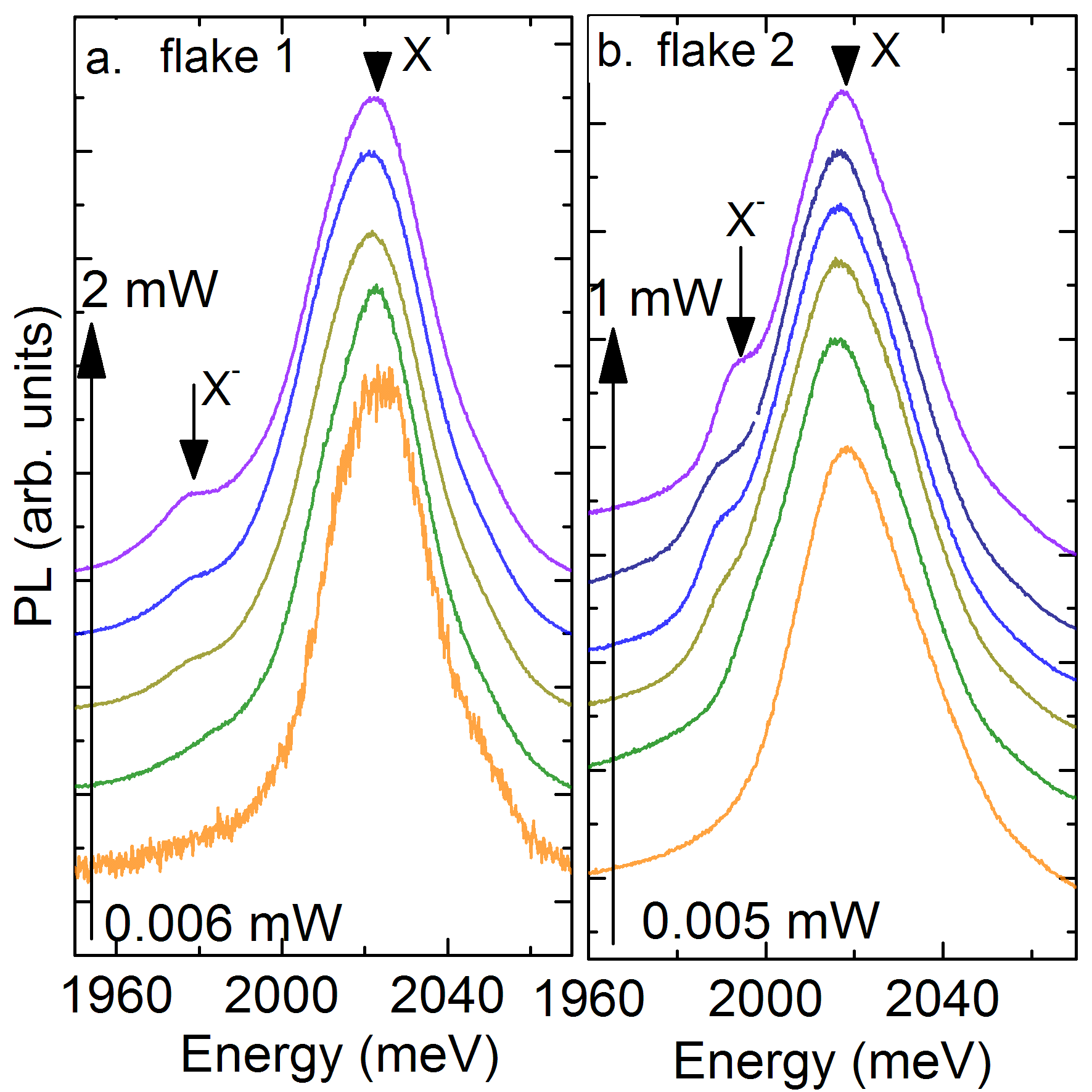}
\end{center}
\caption{(color online) Typical $\mu$PL spectra measured as a function of the 532nm excitation power at 4K. (a),(b)
results obtained for two different flakes. Spectra are offset vertically for clarity}\label{Fig2}
\end{figure}

Perhaps the most striking feature of the $\mu$PL spectra is that we observe a second emission line on the low energy
side of the neutral exciton emission at around $1980$ meV. This line, which becomes much stronger as the excitation
power is increased, is observed in all the flakes we have measured (see \emph{e.g.} data for flake 2 in
Fig~\ref{Fig2}(b). We will see below that the behavior of the low energy line has all characteristics of a trion; with
increasing excitation power, the intensity increases linearly (bi-exciton has quadratic dependence), the binding energy
varies linearly, while the ratio of the trion and neutral exciton emission intensity is not constant.

\begin{figure}[]
\begin{center}
\includegraphics[width= 8.5cm]{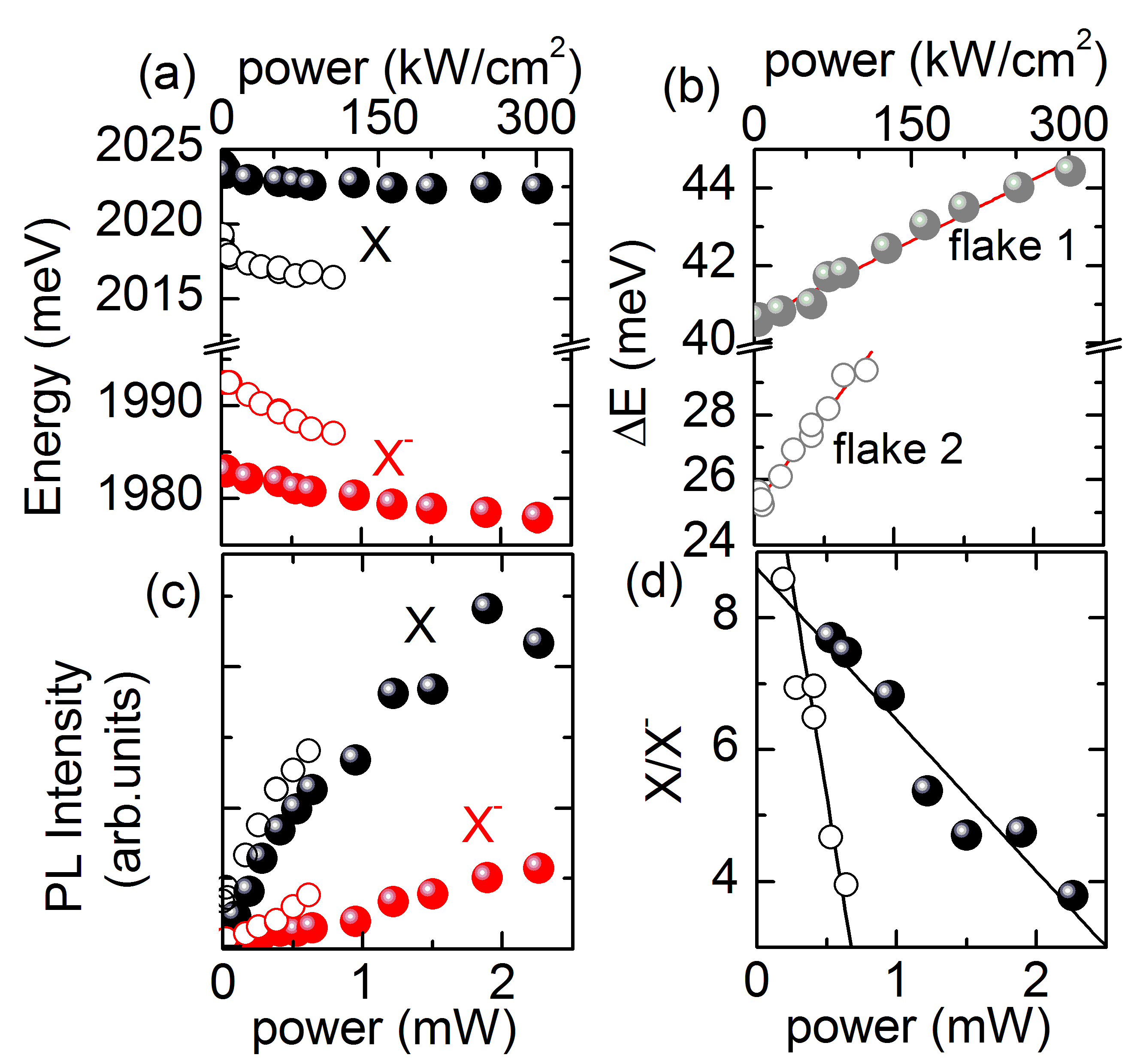}
\end{center}
\caption{(color online) (a) The emission energy of the charged and the neutral exciton  and their difference (b) as a
function of excitation power.  (c) integrated intensity of the charged and the neutral exciton and their ratio (d) as a
function of excitation power. Data is shown for flake 1 (closed symbols) and flake 2 (open symbols). The lines show the
results of a linear fit.}\label{Fig3}
\end{figure}

The neutral exciton is the ground state of a charge neutral system and trions are only formed in the presence of excess
charge which therefore directly controls the intensity of the trion emission. Thus, trion emission is usually not
observed in exfoliated samples of MoS$_{2}$ and WS$_{2}$ unless a gate is
used.~\cite{Eda11,Mak10,Splendiani10,Gutierrez13} Gated samples can exhibit an additional line $20-35$~meV below the
excitonic line which is attributed to the emission from charged excitons~(X$^{\pm}$).~\cite{Mak13,Ross13} Simply by
varying the applied gate voltage it is possible to tune the ratio between neutral and charged exciton emission. Hence,
in our ungated samples, we attribute the low energy line to emission from charged excitons (trions) which we presume to
be negatively charged given the n-type nature of the bulk crystals. The observation of a $X^-$ requires the presence of
excess electrons in the conduction band. In our truly 2D samples internal electric fields cannot spatially separate
oppositely charged particles which makes it almost impossible to create excess charge from photo-created electron-hole
pairs. However, at low temperature Hall data on the bulk crystals shows that the carriers are frozen out onto the donor
levels. Nevertheless, laser illumination, in addition to creating electron-hole pairs, is expected to dynamically
photo-ionize carriers trapped on the donors at low temperature, creating a non equilibrium excess electron density in
the conduction band. Therefore, we expect that the intensity ratio between neutral and charged exciton can simply be
tuned by the varying the power of the laser. To see this more in detail we have analyzed the energy and the intensity
of the emission of both X and X$^{-}$ as a function of the excitation power. The data has been fitted using two
gaussian functions to extract the intensity and position of the two lines. The analysis, performed on flakes 1 and 2,
is presented in Fig~\ref{Fig3}(a),(c).

Fig~\ref{Fig3}(a) shows the power dependence of emission energy of the charged and the neutral exciton. With increasing
excitation power the neutral and charged exciton emission red shifts towards lower energies with a larger change for
the trion. The difference $\Delta E$, between the emission energy of X and X$^{-}$, is the dissociation energy of the
trion which is presented in Fig~\ref{Fig3}(b). The trion dissociation energy grows linearly with the excitation power
suggesting that the density of the carriers also grows linearly with the laser power: The difference in the energy can
be expresses as sum of the binding energy and Fermi energy ($E_{X^{-}}+E_{F}$).~\cite{Huard00} Here binding energy is
defined as the dissociation energy in the limit of infinitively small doping ($E_{F}=0$) where $E_{X^{-}}$ is the
energy needed to promote one electron from the trion to the bottom of the conduction band. The Fermi energy increases
with increasing power of illumination (photon flux) due to the photo-ionization of donors. The trion dissociation
energy (Fermi energy) increases more rapidly with laser power in flake 2 suggesting a larger concentration of donors in
this flake. As the carrier concentration increases, the trion dissociation energy increases and we can tune it over
approximately 4~meV for both flakes. Such a behavior is also observed in II-VI QWs where the interplay between exciton
and charged exciton was achieved using external illumination of the sample.~\cite{Kossacki99,Kossacki04} The enhanced
trion binding energy, compared to standard semiconductors, is linked to the true 2D character of the dichalcogenides
and the larger effective mass.~\cite{Ramasubramaniam12}

The calculated integrated intensity of the emission of the charged and neutral exciton, is presented in
Fig.~\ref{Fig3}(c). For both lines the emission intensity increases with increasing power, however, the increase is not
the same for the neutral and the charged exciton. To illustrate this the exciton/trion intensity ratio is plotted as a
function of laser power in Fig.~\ref{Fig3}(d). For both flakes the ratio of the integrated intensities decreases with
the excitation power by a factor of more than $2$. This is due to increase probability of the creation of the trion
with increasing the density of the photo-ionized carriers. However, as the above gap green laser illumination also
creates electron hole pairs, the intensity of the neutral exciton emission also increases with laser power. The
decrease of the intensity ratio is much faster for the second flake as can also seen directly in the spectra in
Fig~\ref{Fig2} (b), where the trion line is rapidly becoming stronger with increasing excitation power. Again, the
increased sensitivity to laser power is consistent with a larger density of donor impurities in flake 2. Note, that for
both flakes we remain in the linear regime for the trion emission (the exciton emission shows some signs of
saturation), so that even at maximum laser power a significant proportion of the donors are still neutral.

\begin{figure}[]
\begin{center}
\includegraphics[width= 8.5cm]{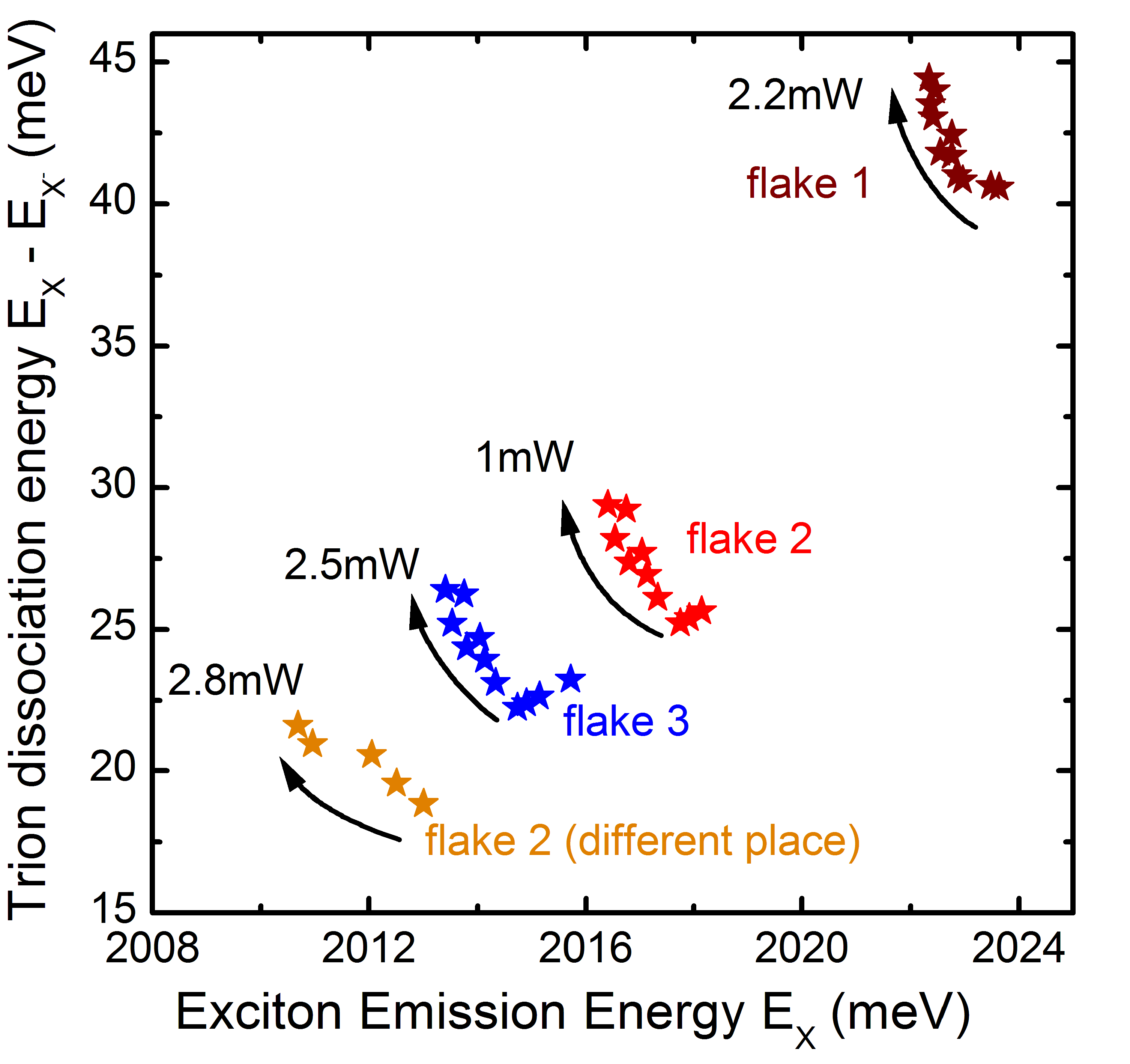}
\end{center}
\caption{(color online) The trion dissociation energy ($E_X - E_{X^-}$) as a function of the exciton emission energy
($E_X$) for three different flakes of WS$_2$. For a given flake the exciton emission energy was tuned by varying the
Fermi energy via the excitation power used. The arrows show the direction of increasing power from close to zero to the
maximum indicated. Green excitation (532nm) was used with the exception of the ``flake 2 (different place)'' data which
corresponds to the two color data in Fig.\ref{Fig4}.}\label{Fig3B}
\end{figure}

In Fig.~\ref{Fig3B} we plot the trion dissociation energy ($E_X - E_{X^{-}}$) as a function of the exciton emission
energy ($E_X$) for three different flakes. For a given flake the trion dissociation energy could be tuned over small
range by varying the power of the 532nm excitation. As previously discussed, with increasing power the trion
dissociation energy increases due to the increase in the carrier density (Fermi energy). For all flakes we can tune the
trion dissociation energy over a range of $\simeq 4$~meV. The increase in the trion dissociation energy is accompanied
by a small decrease ($1-2$~meV) of the exciton emission energy which we attribute to an increase in the exciton binding
energy due to the reduced screening by trions which are in a spin singlet state and therefore cannot screen effectively
due to the Pauli principle. It is important to note that there is a considerable increase $\simeq 20$~meV in the trion
dissociation energy when going from flake 3 to flake 1, which is accompanied by a smaller increase $\simeq 10$~meV in
the exciton emission energy. Within the 2D hydrogen model the exciton binding energy is $E_B=4R_y^*$, where $R_y^*$ is
the effective Rydberg, and the trion dissociation energy is $\approx R_y^*/2 = E_B/8$.~\cite{Ross13} From flake to
flake the trion dissociation energy varies in the range $20-40$meV corresponding to a change in exciton binding energy
within the 2D hydrogen model from $160$ to $360$~meV. The observed increase in the exciton emission energy is much
smaller ($\simeq 10$~meV) so that the increase in exciton binding energy when going from flake 3 to flake 1 has to be
compensated by a similar ($\simeq 170$~meV) increase in the band gap of the crystal.

\begin{figure}[]
\begin{center}
\includegraphics[width= 8.5cm]{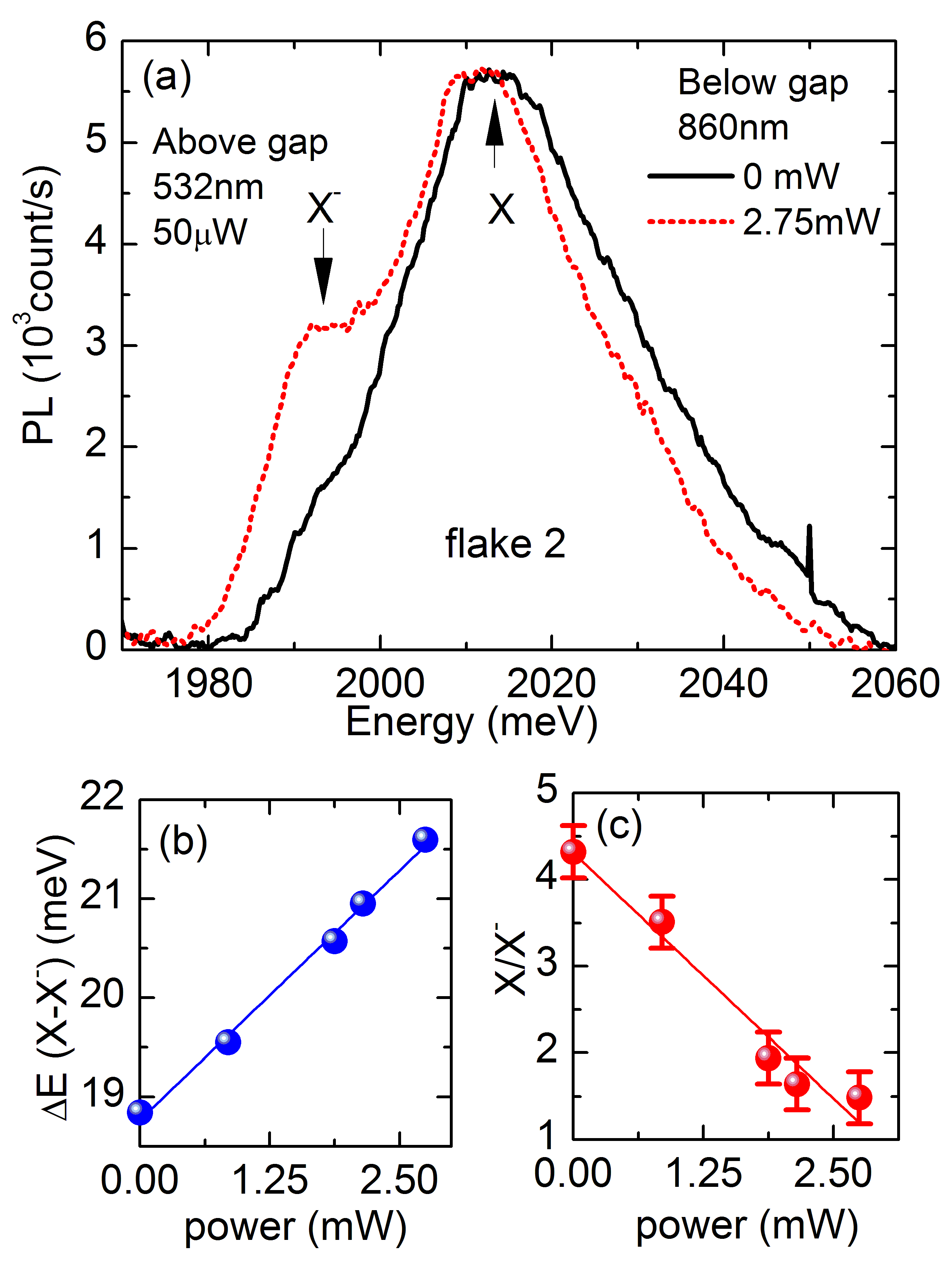}
\end{center}
\caption{(color online) (a) Typical $\mu$PL spectra measured with and without additional below gap illumination. All
spectra are measured with a low constant power of the green (532nm) laser (50$\mu$W).  The difference between emission
energy of the neutral and charged exciton (b) as a function of red (860nm) laser power (below gap excitation). The line
shows the results of a linear fit. (c) The ratio of the integrated intensity of the charged and the neutral exciton as
a function of the red laser power.}\label{Fig4}
\end{figure}

\subsection{Independently tuning the trion emission}

In order to tune the trion and neutral exciton emission independently we have performed $\mu$PL using two color
excitation. A low power (50~$\mu$W) for the above gap green excitation is used in order to generate a constant density
of electron hole pairs together with a small density of photo-ionized electrons. Additional below gap excitation is
provided by a laser centered at 860 nm (1441 meV). Photons with this energy, which is well below the gap, do not
generate electron-hole pairs. However, their energy is largely sufficient to photo-ionize electrons from the donor
level which is a few hundred meV below the indirect conduction band in bulk crystals. Therefore, we expect that the sub
band gap radiation can be used to independently tune the excess density of electrons in the sample.

\begin{figure}[]
\begin{center}
\includegraphics[width= 8.5cm]{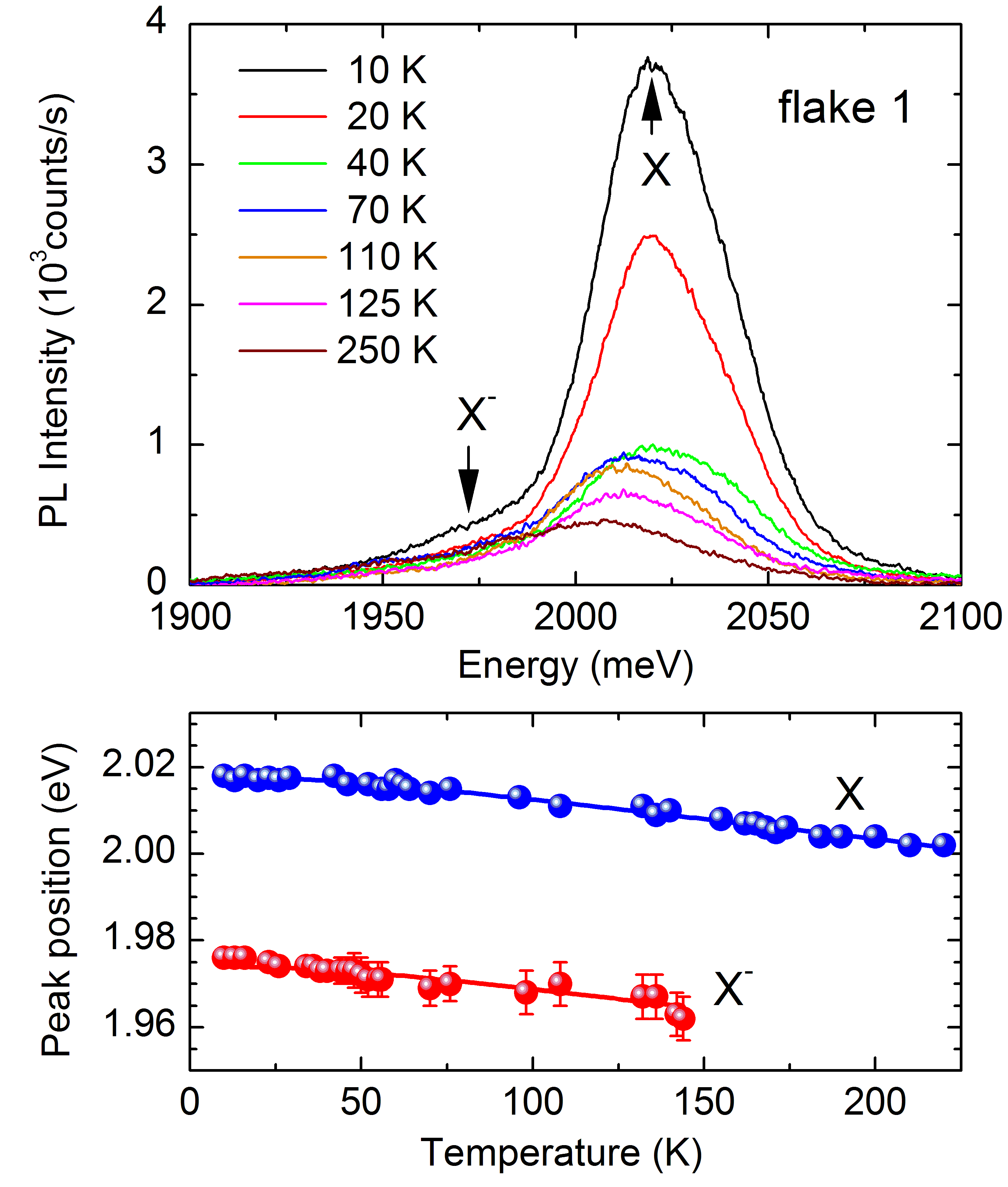}
\end{center}
\caption{(color online) (a) Typical $\mu$PL spectra measured for different temperatures on flake 1. (b) Emission energy
of the neutral and charged exciton as a function of temperature. Solid lines correspond to a fit using
equation~\ref{tempdep} }\label{Fig5}
\end{figure}

Representative two color $\mu PL$ spectra measured on flake 2 are presented Fig~\ref{Fig4}(a). In the absence of below
gap illumination, for the low green excitation power chosen, the trion line is very weak. However, when additionally
illuminated using a laser at 860~nm, the intensity of the trion emission increases while the (integrated) intensity of
the neutral exciton emission decreases. The difference of the emission energy ($\Delta$E) and ratio of the integrated
intensity between neutral and charged exciton determined from the full data set are presented in Fig~\ref{Fig4} (b) and
(c) respectively. We observe that, as for green excitation, the trion dissociation energy $\Delta$E increases linearly
with the power of laser and that the intensity ration between neutral and charged exciton also decreases linearly. This
demonstrates that below band gap excitation can be used to vary the electron density in the conduction band and thus to
independently tune the intensity of the trion emission.

\subsection{Temperature dependence of the band gap}

Finally, the temperature dependance of the emission is presented in Fig~\ref{Fig5}(a). Both exciton lines red shift as
the temperatures increases and the charged exciton emission vanishes around 140~K. The energy of the emission, obtained
by fitting Gaussians, as a function of the temperature is presented in Fig~\ref{Fig5}b. Using the standard expression
for the temperature dependence of a semiconductor band gap~\cite{Donnell91} we can write an expression for the exciton
emission energy as a function of temperature,
\begin{equation}\label{tempdep}
    E(T) = E_{0} - S\langle\hbar\omega\rangle[coth(\frac{\langle\hbar\omega\rangle}{2k_{B}T})-1]
\end{equation}
where $E_{0}$ is the emission energy at zero temperature, $S$ is a dimensionless coupling constant and
$\langle\hbar\omega\rangle$ is the average phonon energy. The best fit to the neutral exciton is obtained with $(E_0 =
2.017$~eV, $S=0.56$ and $\langle\hbar\omega\rangle \simeq 10.4$~meV. The charged exciton is well fitted using the same
parameters except for the zero temperature emission energy $E_0 = 1.973$. Similar values for the average phonon energy
and coupling constant have been obtained for tungsten diselenide.~\cite{Ross13}.

\section{Conclusion}

Using $\mu$PL measurements, we have observed the emission from charged and neutral exciton in an \emph{ungated} single
layer of WS$_{2}$. The trion emission is closely linked to the n-type nature of our crystals. Using two color $\mu$PL
with above and below band gap illumination we can independently tune the trion/exciton intensity ratio. The below band
gap excitations tunes the excess electron density in the conduction band via the dynamic photo-ionization of neutral
donors. Since the photo-ionization threshold will be similar to the donor binding energy $\simeq 200$meV this provides
a possible method for the optical detection of far infrared radiation.

\begin{acknowledgments}
AAM acknowledges financial support from the French foreign ministry. This work was partially supported by NEXTVALLEY
and by ANR project milliPICS.
\end{acknowledgments}


%

\end{document}